\documentclass[usenatbib,useAMS]{mnras}
\usepackage{tablefootnote}
\usepackage{multirow}
\usepackage{xcolor}
\usepackage{hyperref}
\usepackage{graphicx}	
\usepackage{amsmath}	
\usepackage{amssymb}	
\usepackage{multicol}        
\usepackage{bm}		
\usepackage{url}
\usepackage[normalem]{ulem} 
\usepackage{caption}      
\usepackage{longtable}

\title[]{The relationship between FR0 radio galaxies and GPS sources}

\author[A.~Mikhailov and Yu. Sotnikova]{
A.~Mikhailov,$^1$\thanks{E-mail: mag10629@yandex,ru}
and
Yu. Sotnikova,$^1$
\\
$^{1}$Special Astrophysical Observatory of RAS, Nizhny Arkhyz 369167, Russia\\
}

\date{6th CSS/GPS sources workshop proceedings, held in May 10-14, 2021 in Torun, Poland.}
\pubyear{2021}

\begin{document}
\label{firstpage}
\pagerange{\pageref{firstpage}--\pageref{lastpage}}
\maketitle

\begin{abstract}
We present the results of our study of the relationship between FR0 radio galaxies and GPS sources. Quasi-simultaneous radio spectra of 34 FR0s were obtained at 2.25--22.3 GHz with the radio telescope RATAN-600 in 2020--2021 during \mbox{2--6}~epochs. Most FR0s have flat radio spectra, but we found many spectra with a peaked shape. Due to this fact and the compact nature of FR0s, we suggest their possible relationship with CSS/GPS radio sources. We analyzed broadband radio spectra of the 34 FR0s using the RATAN-600 measurements and available literature data. There are 14 FR0 objects which can be CSS/GPS radio source candidates. Most FR0s have broader radio spectra than those of genuine GPS sources, with ${\rm FWHM} >2$ like in blazars. Most spectral indices at the frequencies below and above the peak do not correspond to the values typical of canonical GPS sources. We classified 3 FR0s as low-power GPS sources according to the canonical criteria. The key issue is the variability properties of FR0s. Some FR0s demonstrate a variability level of up to 25\% on a time scale of one year according to the \mbox{RATAN-600} measurements. The flare phenomena in FR0 objects can imply a relationship between them and blazars.
\end{abstract}


\section{Introduction}\label{sec0}

Recent studies have shown that a majority of radio-loud AGNs in the local Universe are relative low power ($\sim 10^{38-40}$ erg/s) and compact, co-called FR0 radio galaxies. In contrast to classical high-power FRI and FRII radio galaxies, FR0s do not have extended radio components, although their radio core and host galaxy properties are simular to FRIs \citep{2018A&A...609A...1B}. The nature of FR0 radio galaxies and their relationship with other classes of radio sources remains poorly understood. The scarcity of broadband radio data and the lack of quasi-simultaneous measurements in the centimeter range were a reason to start the investigation of the radio properties of FR0s with RATAN-600. In this paper we study a possible relationship between FR0 radio galaxies and GPS sources.

\section{The sample and observations}\label{sec1}
 
The observing program for studying radio properties of FR0 radio galaxies has been carried out with the radio telescope RATAN-600 since February 2020. The sample contains 34~objects from FR0CAT \citep{2018A&A...609A...1B} with flux densities $S_{1.4\,{\rm GHz}}>30$ mJy at declinations \mbox{$-08^{\circ}<{\rm Dec}<+47^{\circ}$}. RATAN-600 operates in the transit mode, the spectrum is measured during the horizontal passage of the source across the meridian \citep{1993IAPM...35....7P}. We used the Northern Sector of RATAN-600 and measured instant spectra of objects at six frequencies: $1.28, 2.25, 4,7, 8.2, 11.2$, and $22.3$ GHz quasi-simultaneously. A detailed description of the receivers and antenna parametersare presented in \cite{2019AstBu..74..348S}. To increase the signal-to-noise ratio, we averaged the records of the source passage over a timescale of 7--10 days. The observed data were processed using the Flexible Astronomical Data Processing System (FADPS), developed by \citet{1997ASPC..125...46V} for the broadband RATAN-600 continuum radiometers, and the automated data reduction system \citep{2016AstBu..71..496U}. Thus, we obtained the quasi-simultaneous spectrum for each object in the sample 2--6 times from February 2020 to April 2021. 

\section{Radio properties}\label{sec2}

Radio properties for FR0s from the sample are described in detail in \cite{2021ARep...65..233M}. All the objects are detectable at $4.7$ GHz and their radio luminosities at this frequency are in the range of $4\times10^{38}-4\times10^{40}$ erg/s with a mean value of $\sim\!5\times10^{39}$ erg/s. We revealed that radio loudness $RL$ for many of the objects is close to the boundary value $RL=10$ used to separate radio loud and radio quiet objects.

FR0 radio galaxies usually have flat spectra. We adopted flux density $S\sim\nu^{\alpha}$. Spectral indices $\alpha$ were calculated according to our quasi-simultaneous measurements, and, in general, $|\alpha|<0.5$ for the RATAN-600 frequencies. We note that at 4.7--8.2 GHz the mean spectral index is close to zero. Spectra become steeper at 8.2--11.2 GHz. Thus, FR0s are often characterized by the convex shape of the spectrum. Our quasi-simultaneous spectra have a peaked shape for 40--50\% of the sources in the sample.

\cite{2019MNRAS.482.2294B} observed $18$ FR0s with the VLA and found that many objects were still unresolved at a resolution of $\sim\!0.\!^{''}3$, which corresponds to a linear size of \mbox{100--300}~pc. Features related to extended emission were revealed for only 4 out of 18 objects. The compactness of FR0s is demonstrated by the core-dominance parameter $R$ calculated as a ratio of the flux density at $7.5$ GHz to the flux density at $1.4$~GHz: $R=S_{7.5}/S_{1.4}$. We estimated $R$ using the flux density at $8.2$ GHz, which is close to $7.5$ GHz, and taking into account that our quasi-simultaneous spectra of FR0s are flat in the range of $\sim\!5$--$8$ GHz. In general, our estimates are consistent with the findings of \cite{2019MNRAS.482.2294B} that $\log R \sim 0$ for FR0s.

\section{FR0s with peaked spectra}\label{sec3}

We constructed the continuum radio spectra of FR0 radio galaxies using the RATAN-600 measurements and available literature data, including those from LOFAR and VLASS. We note that the RATAN-600 data allows us to significantly supplement the measurements in the centimeter band \citep{2021ARep...65..233M}. As we mentioned above, many of our objects have peaked quasi-simultaneous spectra. It remains true when we consider available broadband data. About half of the objects have convex spectra with a peak, thereby hinting at a relationship between FR0s and GPS sources.

Criteria for classifying a radio source as a classical GPS source are the following: $\alpha_{\rm below}=+0.5$, $\alpha_{\rm above}=-0.7$, and ${\rm FWHM}\approx1.2$ \citep{1991ApJ...380...66O,1997A&A...321..105D}. We fitted convex-shaped continuum radio spectra of FR0s with a log parabola to determine the peak frequency and estimated the spectral indices below and above the peak. The results are presented in Table~\ref{gps}. The absolute values of the indices differ from the classical for most of the objects. We marked with an asterisk the objects where spectral indices are in good agreement with the classical criteria. Many objects have ${\rm FWHM}>2$ over decades of frequency \citep{2021ARep...65..233M}, thus FR0s have spectra with less spectral curvature than genuine GPS sources. This fact was earlier noted by \cite{2019A&A...631A.176C}. The fourth column in Table~\ref{gps} contains the peak frequencies estimated based on the compiled radio data.

\begin{table}
\centering
\caption{FR0s with peaked broadband spectra\label{gps}}
\begin{tabular}{cccc}
The source & $\alpha_{\rm below}$ & $\alpha_{\rm above}$  & $\nu_{\rm peak}$, GHz  \\ \hline
J0115+0012* & 0.45  & -0.79 & 0.48  \\
J0906+4124  & 0.32  & -0.05 & 11(?) \\
J0909+1928* & 0.45  & -0.79 & 4.63  \\
J0943+3614* & 0.41  & -0.71 & 8.34  \\
J1025+1022  & 0.37  & -0.45 & 0.97  \\
J1057+4056  & 0.23  & -0.64 & 0.84  \\
J1111+2841  & 0.50  & -0.30 & 10.69 \\
J1142+2629  & 0.39  & -0.44 & 2.93  \\
J1205+2031  & 0.03  & -0.39 & 0.37  \\
J1246+1153  & 0.21  & -0.62 & 0.29  \\
J1334+1344  & ?     & -0.68 & 0.39  \\
J1336+0319  & ?     & -0.36 & 1.16  \\
J1604+1744  & 0.21  & -0.38 & 5.67  \\
J1606+1814  & 0.16  & -0.25 & 0.54 \\ \hline
\end{tabular}
\end{table}

\begin{figure*}
\centerline{\includegraphics[width=0.65\textwidth]{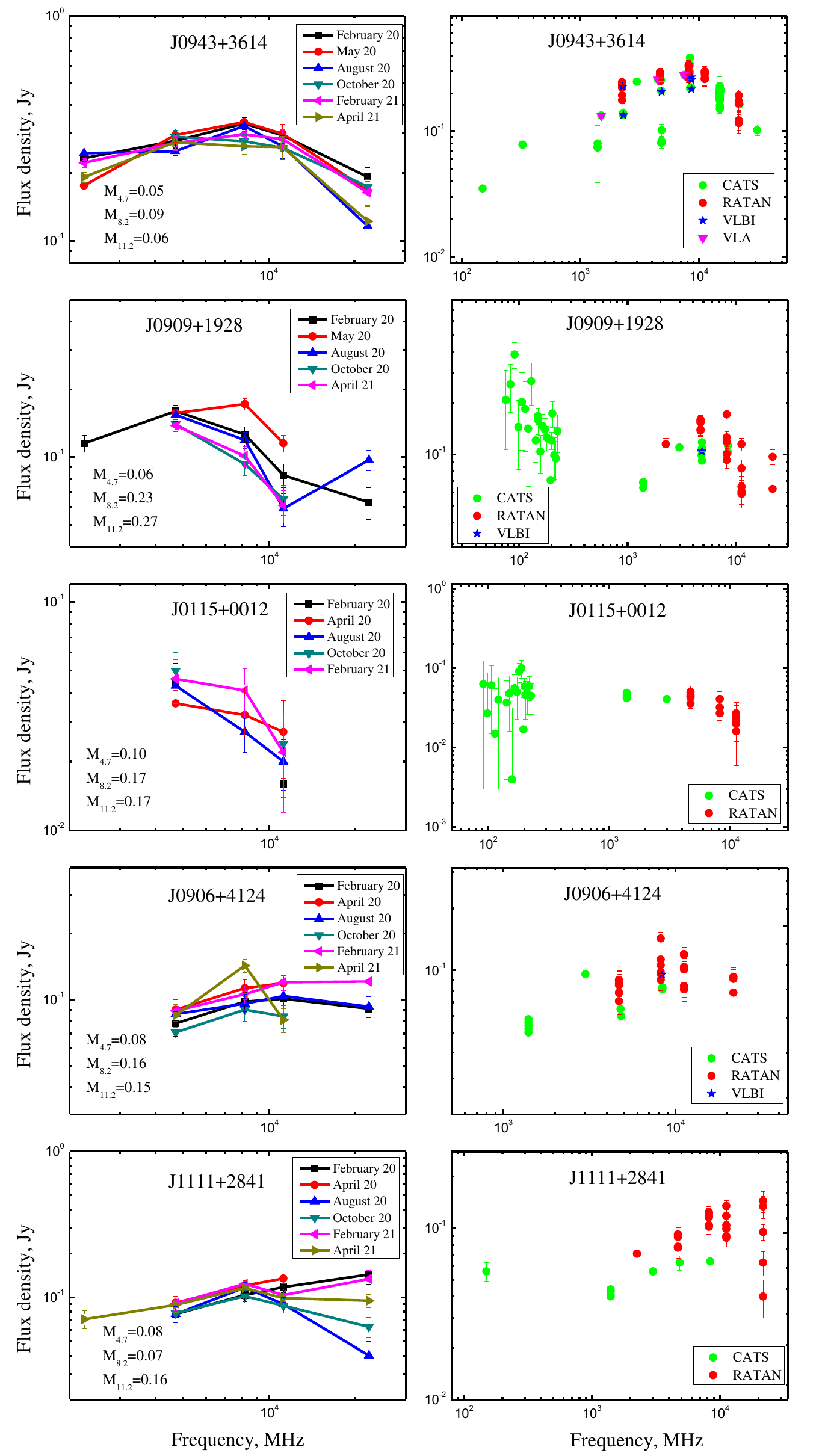}}
\caption{Quasi-simultaneous spectra measured with 
RATAN-600 (left) and broadband radio spectra (right).
\label{fig1}}
\end{figure*}

\section{GPS candidates}\label{sec4}

The spectrum of a radio source can have a peak temporarily due to its flaring activity, therefore such an object can be misclassified as a GPS source (the so-called blazar contamination problem). Long-term monitoring and quasi-simultaneous measurements are crucial for resolving the problem of GPS source classification. To tackle this question in our case, we considered quasi-simultaneous measurements obtained with RATAN-600 and available literature data, which were searched for using the CATS database \citep{1997BaltA...6..275V,2005BSAO...58..118V}. In Fig.~\ref{fig1} we present quasi-simultaneous (left) and broadband (right) spectra for the objects that can be considered GPS sources according to the classical criteria, along with examples of blazar-like objects.

The variability of the sources during the period of \mbox{RATAN-600} observations is characterized by the modulation index $M_{\nu}$ calculated at frequencies of 4.7, 8.2, and 11.2 GHz and presented in Fig.~\ref{fig1}.

Quasi-simultaneous spectra of J0943+3614 have had constant shape during 14 months of RATAN observations. The variability is not bigger than 10\%. According to available data, the peak frequency is 8.3~GHz and the spectral indices correspond to the classical criteria. Thus, J0943+3614 can be classified as an HFP source. We note that the VLBI image of the source has unresolved core-type morphology at 8.3 GHz \citep{2018ApJ...863..155C}.

The shape of the quasi-simultaneous spectrum of J0909+1928 is constant, although this source demonstrates variability of up to 25\%. It can be classified as a GPS source with a peak frequency close to 5 GHz. Its VLBI image has core-jet morphology at 4.8 GHz \citep{2018ApJ...863..155C}.

The source J0115+0012 has also had a constant spectral shape during RATAN observations with variability not bigger than 20\%. The peak frequency is close to 0.5 GHz.

Quasi-simultaneous spectra of the source J0906+4124 do not have a constant shape, the peak frequency changes. This source has core-jet morphology at 8.3 GHz \citep{2018ApJ...863..155C}. The object is a blazar according to \cite{2013MNRAS.430.2464M}.

The source J1111+2841 is quite variable at 22.3 GHz. The peak frequency is not less than 8 GHz, the current estimate is 10.7 GHz. This source has an inverted spectrum over a wide frequency range
and needs further monitoring.
 
\section{Summary}\label{sec5}

We found that almost half of the FR0 objects in the  sample have a peak in their spectra. We conclude that at least 3~objects can be classified as low-power classical GPS sources. Thus, we estimate the fraction of GPS radio sources among FR0s is approximately 10\%. However, FR0s have spectra with less spectral curvature than genuine GPS sources. Some objects show variability of up to 25\% and can be blazars (for example, J0906+4124 and J1111+2841). Thus, the original catalogue of FR0 radio galaxies is heterogeneous and contains low-power AGNs of different types. Further study of the nature of FR0s and their relationship with different AGNs types is necessary.

\section*{Acknowledgments}

The observations with RATAN-600 scientific facility are supported by the Ministry of Science and Higher Education of the Russian Federation. This research has made use of the CATS database, operated at SAO RAS, Russia. 

\bibliographystyle{mnras}
\bibliography{FR0}

\label{lastpage}
\end{document}